# Time-resolved and spectral-resolved optical imaging to study brain hemodynamics in songbirds

Stéphane Mottin[a], Bruno Montcel*[b], Hugues Guillet de Chatellus[c], Stéphane Ramstein[a], Clémentine Vignal[d], Nicolas Mathevon[d].

[a]CNRS; Université de Lyon; Université de St-Etienne, UMR5516, Saint-Etienne, France;
[b]Université Lyon1; CREATIS; CNRS UMR5220; INSERM U1044; Université Lyon1; INSA Lyon, France;
[c]CNRS; Université de Grenoble; UMR5588, St Martin d'Hères, France;
[d]Université de St-Etienne, CNRS, ENES/CNPS UMR8195, Saint-Etienne, France;

## ABSTRACT

Contrary to the intense debate about brain oxygen dynamics and its uncoupling in mammals, very little is known in birds. In zebra finches, picosecond optical tomography (POT) with a white laser and a streak camera can measure *in vivo* oxy-hemoglobin ($HbO_2$) and deoxy-hemoglobin (Hb) concentration changes following physiological stimulation (familiar calls and songs). POT demonstrated sufficient sub-micromolar sensitivity to resolve the fast changes in hippocampus and auditory forebrain areas with 250 µm resolution. The time-course is composed of (i) an early 2s-long event with a significant decrease in Hb and $HbO_2$, respectively -0.7 µMoles/L and -0.9 µMoles/L (ii) a subsequent increase in blood oxygen availability with a plateau of $HbO_2$ (+0.3µMoles/L) and (iii) pronounced vasodilatation events immediately following the end of the stimulus. One of the findings of our work is the direct link between the blood oxygen level-dependent (BOLD) signals previously published in birds and our results. Furthermore, the early vasoconstriction event and post-stimulus ringing seem to be more pronounced in birds than in mammals. These results in bird, a tachymetabolic vertebrate with a long lifespan, can potentially yield new insights for example in brain aging.

## 1. INTRODUCTION

The reliability of optical measurements of changes in the concentration of hemoglobin in tissues has been a challenge for years. Because of its non-invasive nature, transcranial optical cerebral oximetry has become a source of quantitative or semi-quantitative information about brain oxygenation, cerebral blood flow and volume. But continuing technical controversies about signal derivation, accuracy, precision, and quantitative ability have limited its application. However time-resolved techniques have opened up the way to promising methods [1-7]. As part of our broader effort to develop a non-invasive neuro-method and to improve quantitative measurement of absorbing chromophores into scattering brain tissues, we worked on a time-domain based device. Using a white light super-continuum or "white laser", we combined POT with near-infrared spectroscopy (spectral-POT) [2] and a POT with contact free spatial imaging (spatial-POT) [8, 9]. In the near-infrared spectral window 650 to 850 nm, the non-monotonic behavior of the absorption spectrum of Hb provides reliable "molecular fingerprints" [2]. Furthermore the optical signals are integrated into a selected picosecond time-of-flight window specifically defined so as to probe only the targeted deep brain structures [2]. This system allows us to monitor in vivo and quantify, for the first time, an evoked brain hemodynamic response to acoustic stimuli with sub-micromolar sensitivity and sub-millimeter spatial resolution in small songbirds which have become an important focus of interest for comparative neuroscience [10].

## 2. MATERIALS AND METHODS

Adult male zebra finches (*Taeniopygia guttata*) served as subjects for the experiments. 4 birds and 5 birds were used respectively for spectral-POT and spatial-POT. Birds were anesthetized with 2% isoflurane under spontaneous breathing conditions. The animal preparation and the spectral-POT setup have been described previously [2]. For spectral POT, the optical fibers were placed directly on the skin. Positions of the input and collecting optical fibers (inter-fibers distance of 5 mm) were chosen in order to probe the auditory regions of the telencephalon (field L, the caudo-medial Nidopallium NCM and the caudo-medial Mesopallium CMM). The animals were kept in a custom-made sound-attenuated box equipped with 2 fixed high-fidelity speakers. After a 1mn baseline period, each bird was subject to a 20s stimulus, followed by 1mn for recovery of baseline. The original auditory signal was a random sequence of songs and calls recorded in the zebra finch aviary, normalized to the same intensity. All experimental procedures were approved by the University's animal care committee. Statistical methods have been previously described [2]. The variation of the time-resolved transmittance spectrum was also fitted to the spectra of HbO2 and Hb known in mammals by classic linear least-squares procedure. The same procedure was applied to calculate the variations in concentration of HbO2 and Hb. These concentration variations can be expressed using an absolute scale ($\mu$Mol) because our time-resolved detection system can measure the mean optical path through the bird's head thanks to the mean arrival time of photons. The spatial POT has also been described previously [8, 9]. We used the same setup and the same laser fiber position as previously described, with the omission of the polychromator and with an imaging system between the head of the animal and the streak camera (figure 1). Intrinsic filtering properties of the imaging setup enable to collect only the photons emerging from a 5 mm-long segment located 5 mm apart from the input fiber. We put a narrow bandwidth filter (IF) (10 nm FWHM) centered at 700 nm, where the difference of absorption between the two hemoglobin species is maximal. The position of the imaged segment on the head of the bird can be controlled by eye by shining an intermediate slit with a He-Ne laser and checking and adjusting the position of its image on the surface of the head. The spatial resolution along the slit is near of 250 µm.

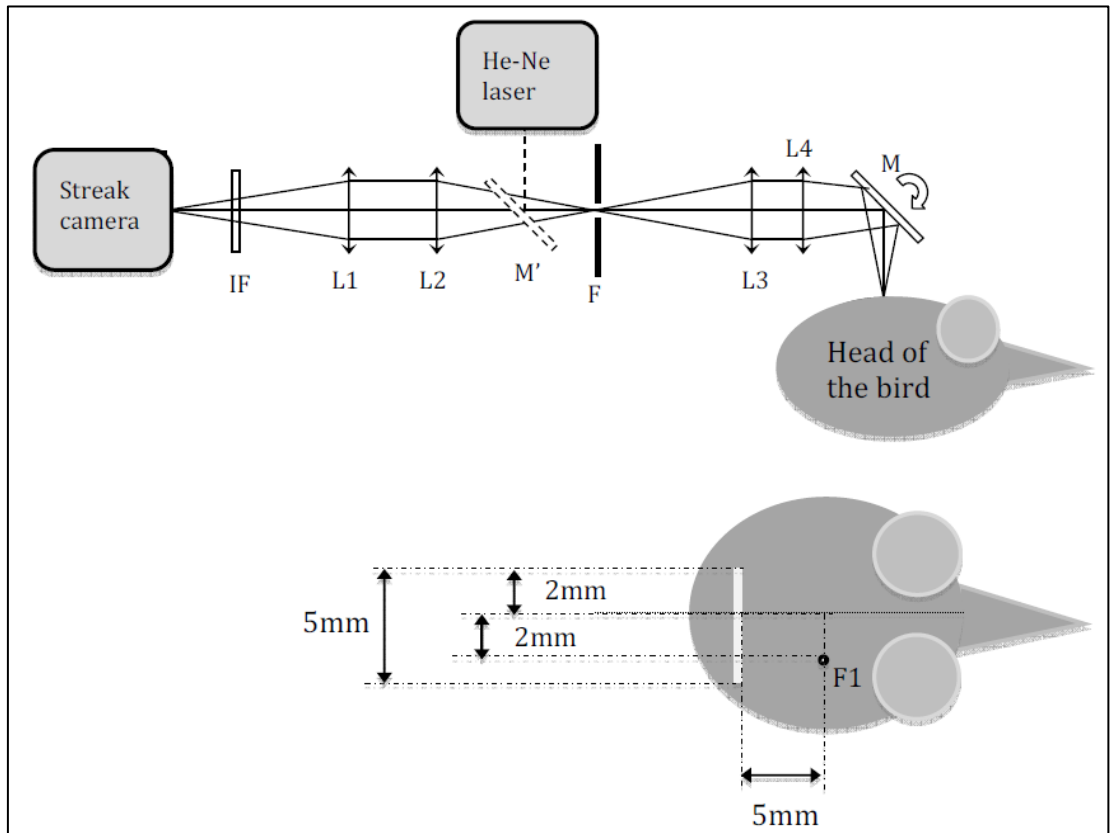

Figure 1 [8, 9]. Experimental set up.

## 3. RESULTS

Figure 2 shows the full time-courses of picosecond time-resolved transmittance measured by spectral-POT (Figure 2a) and by spatial-POT (Figure 2b). The shape of the time-courses and the level of variation of transmittance were equivalent for spatial-POT and spectral-POT. The maximum change in transmittance induced by the auditory stimulus was 1.03. To establish a calibrated functional technique, the acoustic response experiments were carried out under the same conditions as the 7% normoxic hypercapnic experiments [2]. The functional signal under these conditions was found to be equivalent to 10% of the hypercapnic changes. Significant Hb and $HbO_2$ changes were obtained by linear un-mixing and were analyzed with a 0.667 s time resolution. During the 2-seconds following the onset of acoustic stimuli, Hb and $HbO_2$ levels significantly decreased to -0.7 µMoles/L and -0.9 µMoles/L respectively (Figure 3). The $HbO_2$ level then increased significantly (during 12.4s, 100 concentration measurements) to reach a plateau of 0.3 µMoles/L (p=0.015 when compared to the 100 concentrations preceding the stimulus). Immediately after the end of the stimulus, Hb and $HbO_2$ pulses reached +0.7 µMoles/L.

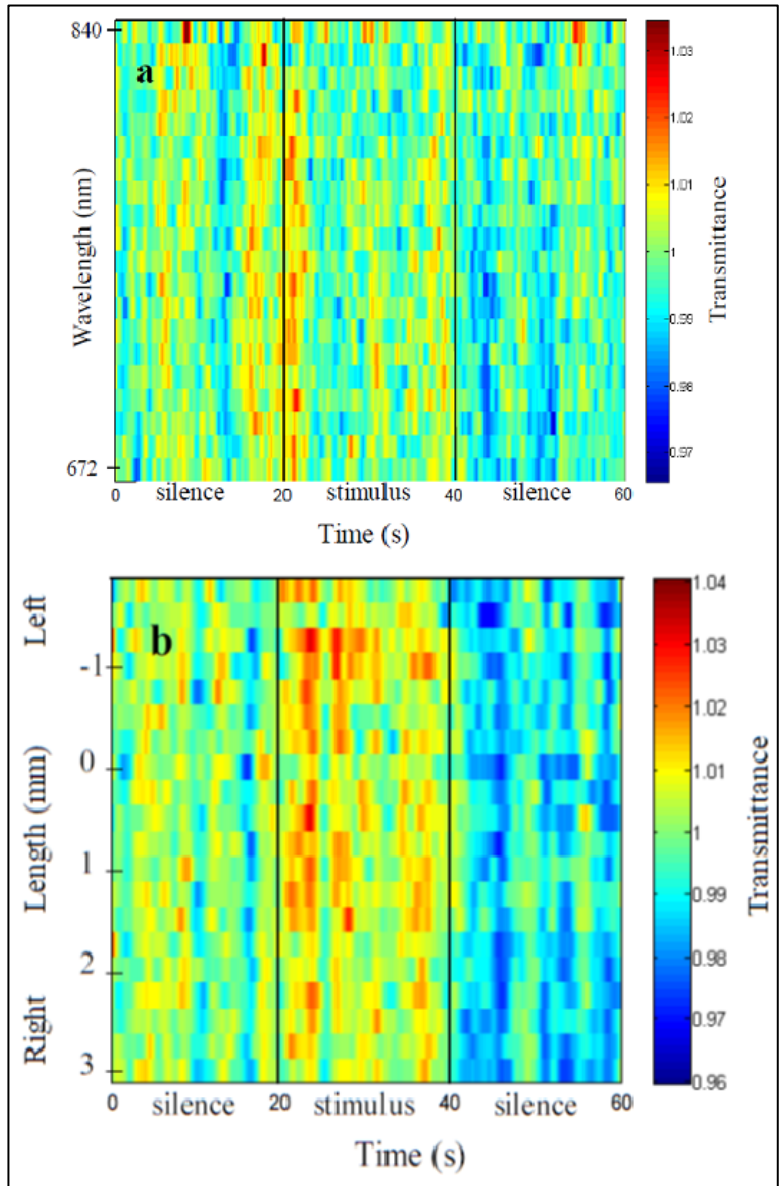

Figure 2 [8, 9]. The full time-course of the picosecond time-resolved transmittance spectra was measured by spectral-POT (a). The near-infrared spectral window is 668-844.4 nm, with 20 spectral windows of 8.83 nm. The time-course of the picosecond time-resolved transmittance for 20 spatial regions of 0.25 mm was imaged by spatial-POT (**b**). The 695-705 nm spectral window was used for spatial-POT. Each point of measurement corresponds to 33ms. For illustration purposes, these results were filtered to get rid of high frequency noise, using a Chebyshev window only along the time axis, 1s for spectral-POT and 2s for spatial-POT respectively. The 0 mm position corresponds to the sagittal midline

Changes were significantly localized (Figures 2b) above the auditory forebrain areas. There was a significant bilateral increase in transmittance when compared to more lateral positions [2.75 and 3 mm]. During the post-stimulus period all areas showed significant decreases in transmittance when compared to the rest period. These results show that re-coupling was less localized than uncoupling. Compared to more lateral positions and within the stimulation period, a significant bilateral increase was observed. In contrast, a significant bilateral decrease was observed within the post-stimulus period. Furthermore the number of Hb and $HbO_2$ pulses was less high for the auditory-hippocampal areas [0.25 to 1.25mm and -0.25 to -1.25mm] than for more lateral positions, showing that re-coupling was faster in these areas (Figures 2b).

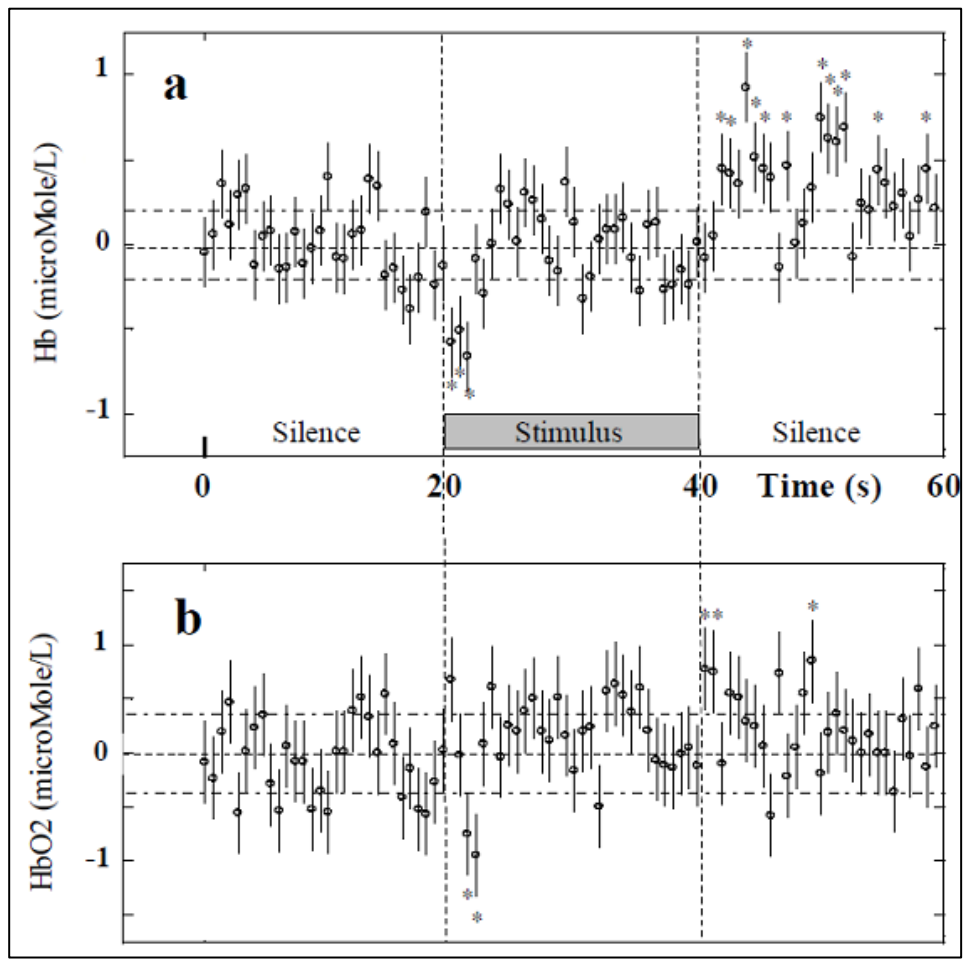

Figure 3 [8, 9]. Hb (**a**) and HbO2 (**b**) concentration changes obtained by linear unmixing of the picosecond time-resolved transmittance spectra. Each point is an average of five concentrations, allowing a time resolution of 0.667s. Bars corresponds to p=0.05 for multiple comparisons (one-way ANOVAs for repeated measures) between periods. The limits of significance of Hb and HbO2 are 0.42 µMoles/L and 0.75 µMoles/L respectively. The asterisk indicates significant difference (p<0.05) from the detection limit.

## 4. DISCUSSION

Our study demonstrates for the first time the changes of blood oxygen in a small songbird during stimulation *in vivo*. The most intense responses to similar stimuli have been observed in NCM and Field L, using other methods [11, 12]. In addition to bilateral responses in these areas, lateralized activation has been suggested to take place [11, 12]. With our less specific stimulation paradigm, we obtained a bilateral response without lateralization (Figures 2b). For the post-stimulus period, re-coupling seems to be more complex than expected because (i) the $HbO_2$ and Hb pulses were less localized than during activation (Figure 1b), (ii) the re-coupling of the activated auditory regions was faster than for other regions, and (iii) the early $HbO_2$ pulse arrived before the Hb pulse (Figure 3). We imaged the sinus sagittalis superior (position 0 mm in Figure 2b) and no significant changes were observed during the activation period. However this result should be considered carefully because distinguishing arterial, capillary and venous compartments is not straightforward in optical neuro-imaging [13].

## 5. CONCLUSION

Our study demonstrates the occurrence of strong reactivity in the cerebral vessels of the bird, an animal with a long lifespan [14]. Several studies [15] indicate that age-related changes in vascular reactivity are important contributing factors to mild cognitive impairment in aging mammals. Contrary to accepted dogma, the role of oxidative stress as a determinant of longevity is still open to question [14, 15]. Our results could thus shed light on this crucial question i.e. the link between brain aging and vascular reactivity.

## 6. ACKNOWLEDGEMENTS

These experiments were supported by the Program 'Emergence' of the Région Rhône-Alpes and the Agence Nationale de la Recherche (Project 'Birds' voices', ANR-06-BLAN-0293-01)